\documentclass[10pt]{article}
\usepackage{OSAmeet}
\usepackage{epsfig}

\def\ket#1{| #1 \rangle}

\begin{document}
\title{Quantum control of population inversion in the presence of spontaneous emission}
\author{S.~G.\ Schirmer, Andrew D.\ Greentree and A.~I.\ Solomon}
\address{Quantum Processes Group, Departments of Applied Maths and Physics and Astronomy,
         The Open University, Milton Keynes, MK7 6AA, United Kingdom}
\email{Tel: +44-1908-652326, Fax: +44-1908-652140, S.G.Schirmer@open.ac.uk}
\HeaderAuthorTitleMtg{Schirmer, Greentree and Solomon}{Quantum control of population inversion}{CQ08}

\begin{abstract} 
The detrimental effect of spontaneous emission on the performance of control schemes 
designed to achieve population inversions between the ground state and a highly excited
atomic state is studied using computer simulations.
\end{abstract}
\ocis{020.0020,140.7242,270.3430}

\noindent                      
We consider a $N$-level quantum system with energy levels $E_n$ and corresponding 
energy eigenstates $\ket{n}$ for $n=1,2,\ldots,N$.  Assuming that the system is 
initially in the ground state $\ket{1}$, our objective is to maximize the population
difference between the ground state and the excited state $\ket{N}$ by applying a 
sequence of simple control pulses.  For simplicity, we shall restrict our attention 
here to control schemes involving only transitions between adjacent energy levels in
the model and control pulses that are resonant with one of the allowed transition 
frequencies.  It will be assumed that the allowed transition frequencies $\mu_n=
E_{n+1}-E_n$ for $n=1,2,\ldots,N-1$ are distinct and sufficiently separated so that 
off-resonant effects can be neglected.

\begin{figure}
\begin{minipage}{3.2in}
   \centerline{\scalebox{.55}{\includegraphics{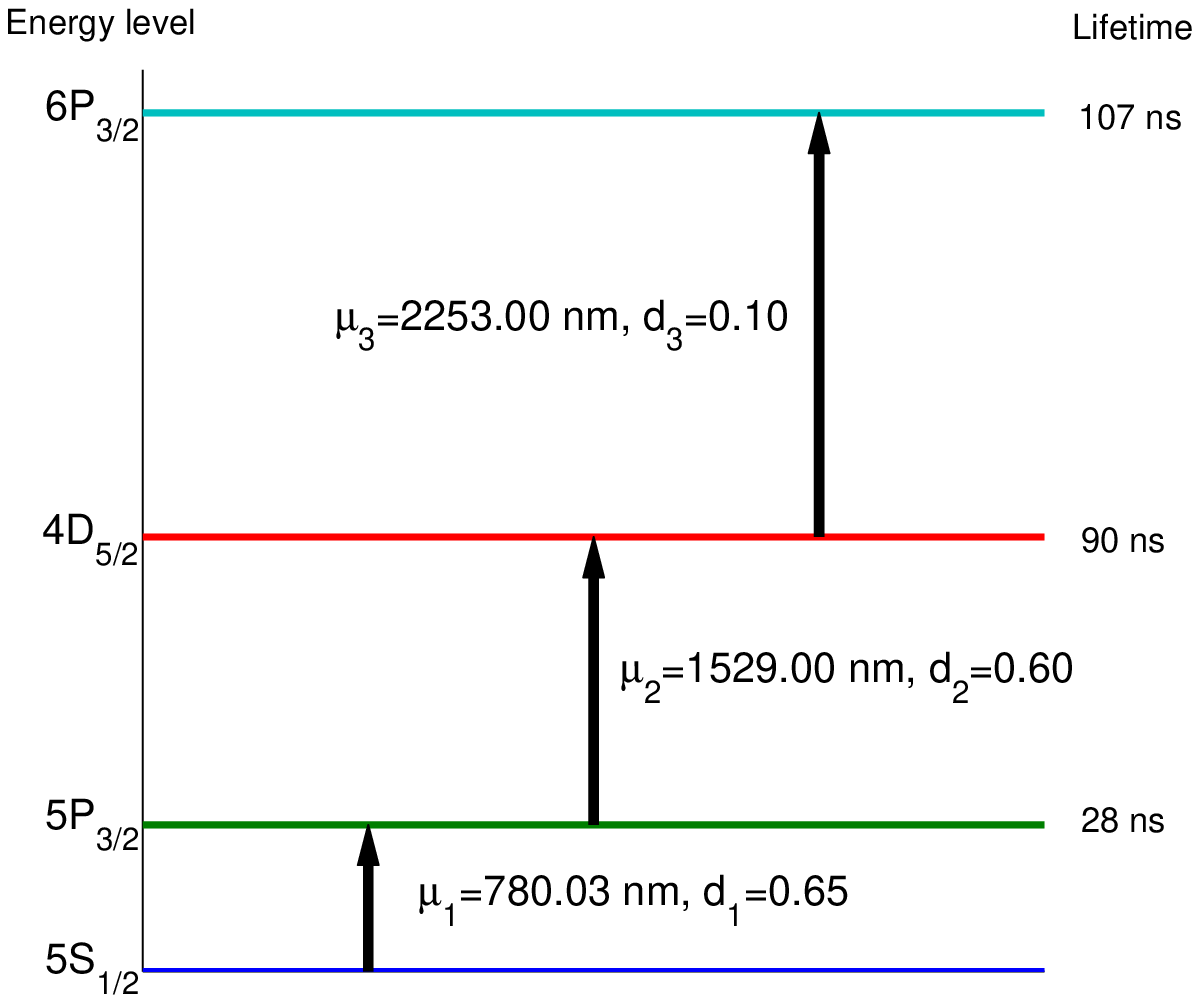}}}
   \caption{Energy-level diagram}\label{Fig:Ru} 
\end{minipage}
\hfill
\begin{minipage}{3.2in}
   \centerline{\scalebox{.5}{\includegraphics{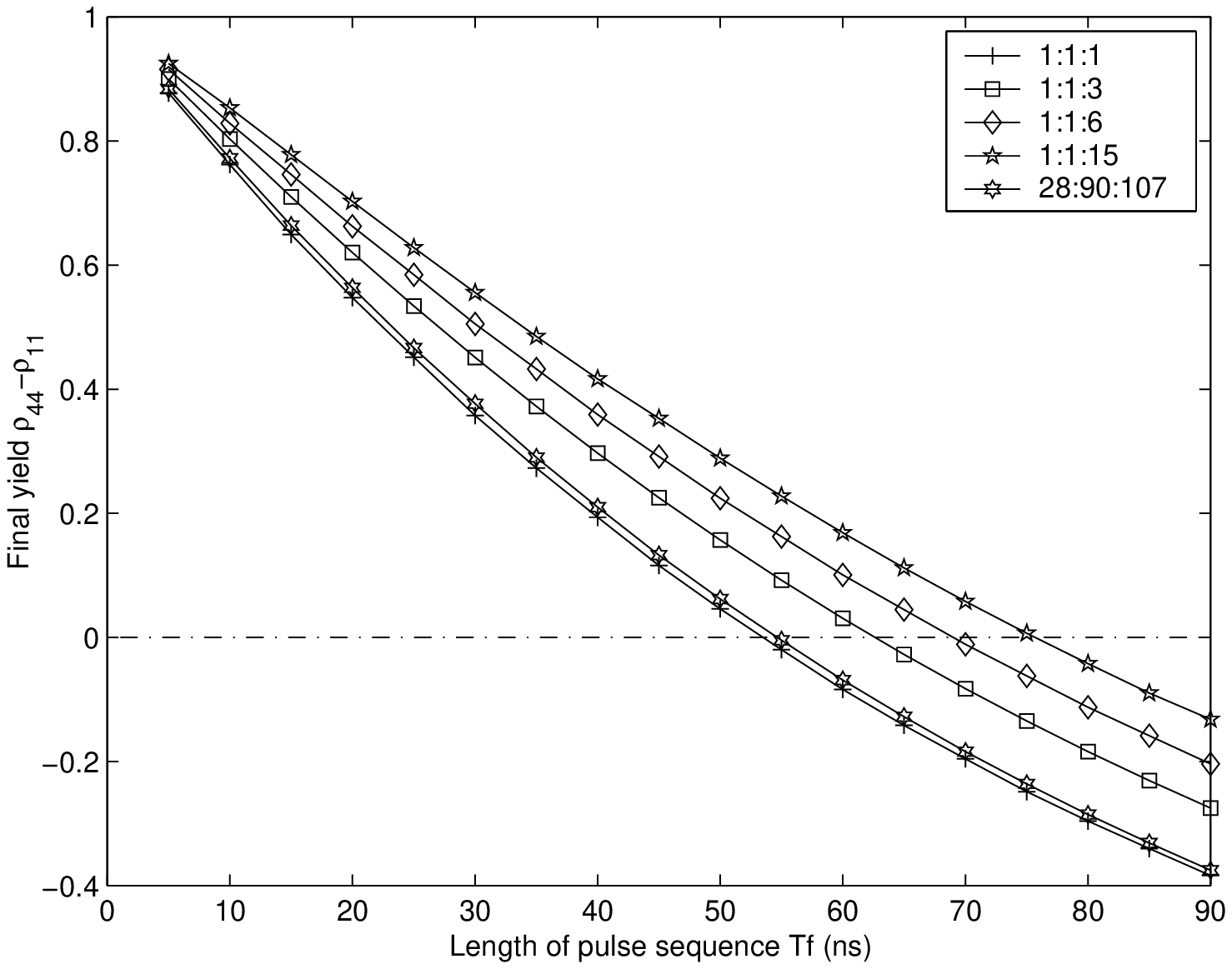}}}
   \caption{Final value of $\rho_{44}-\rho_{11}$ as a function of $T_f$}\label{fig:yield}
\end{minipage}
\end{figure}

If we neglect the finite lifetimes of the excited states and assume that each control 
pulse interacts only with its resonant transition, then it can easily be shown, e.g., 
using Lie group decompositions \cite{vish00}, that a complete population transfer from
the ground state to the highest excited state $\ket{N}$ can be achieved by applying a
sequence of $N-1$ control pulses such that the $n$th pulse is resonant with the $n$th
transition frequency $\mu_n$ and has a total pulse area of $\pi/(2 d_n)$, where $d_n$
is the absorption oscillator strength of the transition $\ket{n}-\ket{n+1}$. The length
and shape of the pulses do not influence the outcome of the control process as long as
the pulse areas remain fixed.  
For a real system, however, the decay of the excited states by spontaneous emission 
reduces the populations of the excited states and ultimately leads to the repopulation 
of the ground state, which is detrimental to achieving population inversion.  Our aim 
is to study the effect of spontaneous emission on the outcome of the control process 
using computer simulations.  Due to space constraints, we only consider a four-level 
model of Rubidium with energy levels and transitions as indicated in Fig.~\ref{Fig:Ru}
as a concrete example.  If dissipative effects are neglected then it can easily be seen
that a complete population transfer from the ground state $\ket{1}=\ket{5S_{1/2}}$ to 
the excited state $\ket{4}=\ket{6P_{3/2}}$ can be achieved by applying a sequence of 
three control pulses of frequencies $\mu_1$, $\mu_2$ and $\mu_3$ and pulse areas 
$\pi/(2d_1)$, $\pi/(2d_2)$ and $\pi/(2d_3)$, respectively, and that the result of the
control process is independent of the shape and length of the control pulses.  If 
dissipative effects such as the finite lifetimes of the excited states are taken into
account, however, then a complete population transfer can usually not be achieved, and
the performance of the control process depends significantly on the shape and length 
of the control pulses employed.

Fig.~\ref{fig:yield} shows the difference between the populations of states $\ket{1}$ 
and $\ket{4}$, $\rho_{44}-\rho_{11}$, at the conclusion of the control process as a 
function of the total control time $T_f$ for various ratios of the pulse lengths, as 
obtained from a computer simulation of the control process.  Note that for very short 
control times the final yield is around 90\% of the theoretical maximum value of one. 
As the total control time increases, the final yield $\rho_{44}-\rho_{11}$ decreases 
since dissipative effects accumulate.  However, the computer simulation suggests that
the final yield also depends considerably on the ratio of the pulse lengths, not just
the total control time $T_f$.  For our four-level Rubidium system, the naive choice of
equal pulse lengths, i.e., pulse ratios $1:1:1$, consistently results in the worst yield,
independent of the total control time $T_f$. Another obvious choice of the pulse lengths 
according to the ratio of the lifetimes of the excited states results only in a marginal
improvement.  However, the final yield and the threshold value of $T_f$ for population
inversion increase considerably as the ratio of the first two pulses versus the last 
pulse increases.  These at first quite startling results can be partly explained by the
vastly different lifetimes of the excited states of Rubidium.  The lifetime of state 
$\ket{2}=\ket{5P_{3/2}}$ is less than a third of that of $\ket{3}=\ket{4D_{5/2}}$ and 
almost one fourth of that of $\ket{4}=\ket{6P_{3/2}}$.  Hence, state $\ket{2}$ is the 
weakest link in the chain and it is therefore imperative to minimize the time the system
spends in state $\ket{2}$ in order to reduce dissipative losses.  However, choosing the 
pulse lengths according to the ratio of the lifetimes of the excited states is not 
sufficient since the amount of time the system spends in state $\ket{n}$ in the control
 scheme under consideration is roughly proportional to $(\Delta t_{n-1}+\Delta t_n)/2$,
not simply $\Delta t_n$.  We see that it is advantageous to choose the pulse lengths 
$\Delta t_n$, $n=1,2,3$, such that $\Delta t_2+\Delta t_3 >3(\Delta t_1 + \Delta t_2)$, 
which explains why increasing the ratio of the pulse length of the third pulse compared
to that of the first two pulses tends to reduce the repopulation of the ground state due
to spontaneous emission and thus improves the performance of the control scheme.  

\begin{figure}
\begin{center}
   \scalebox{.5}{\includegraphics{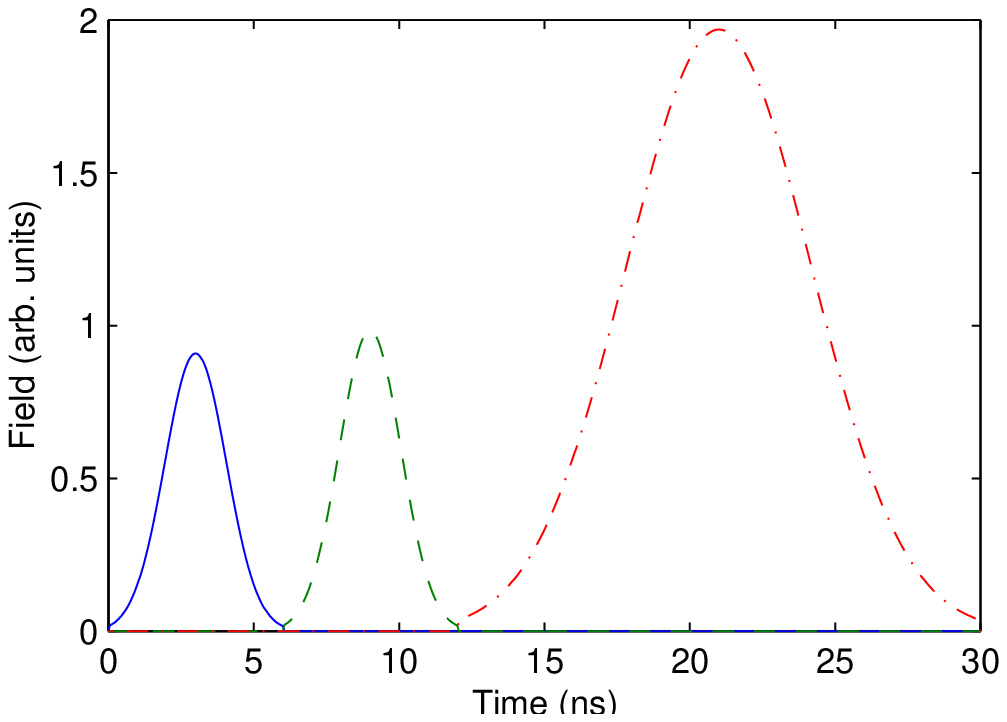}}
   \scalebox{.5}{\includegraphics{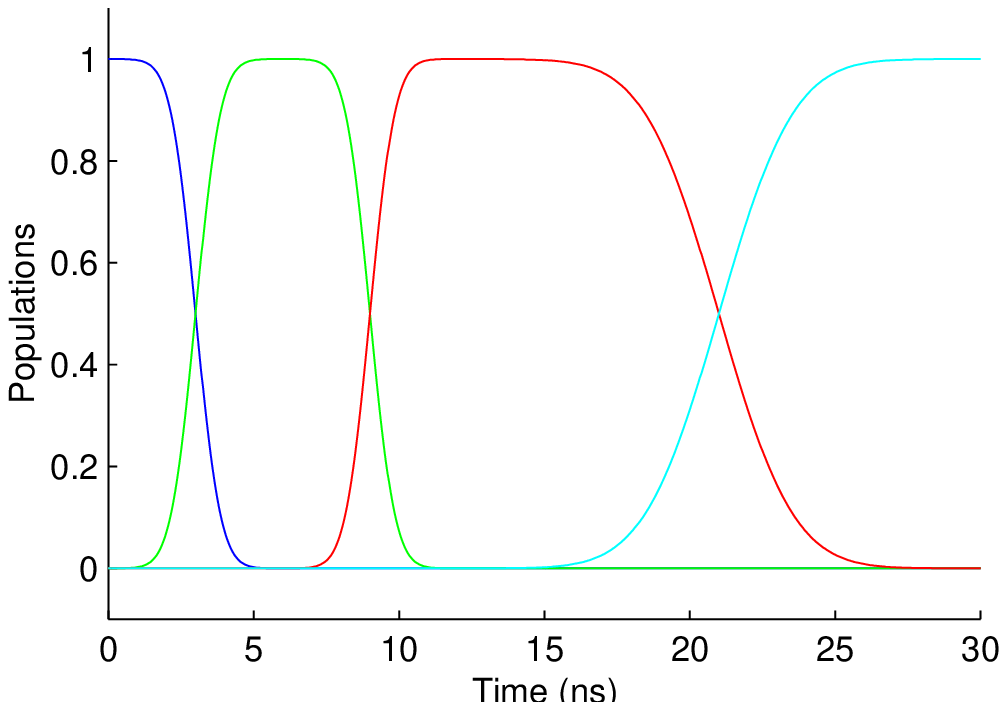}}
   \scalebox{.5}{\includegraphics{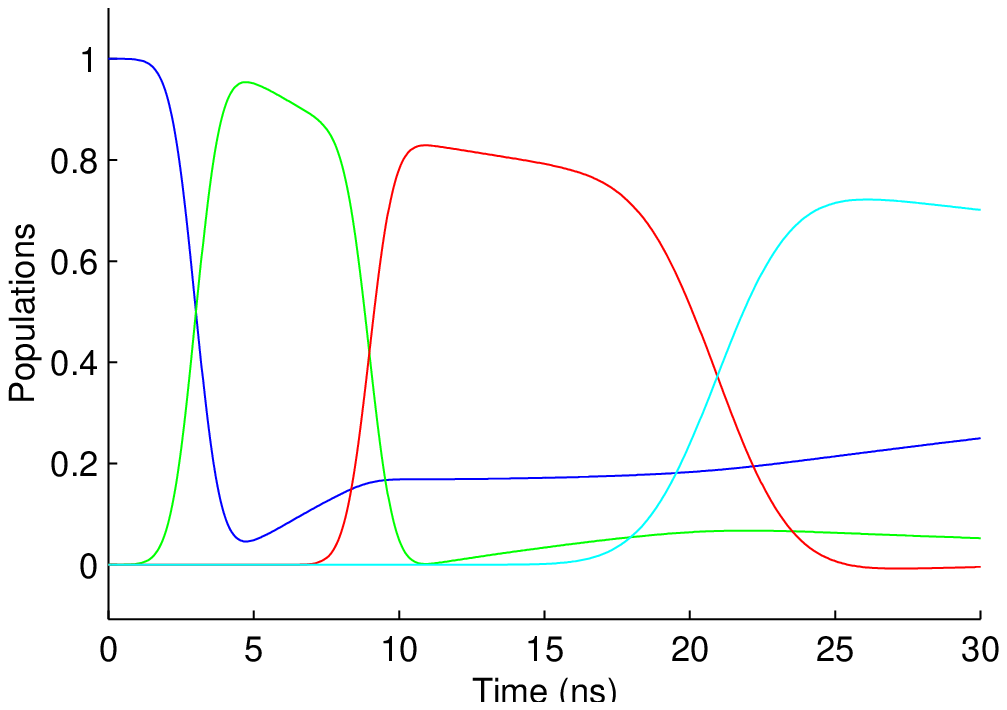}}
\caption{Control pulse sequence (left) and corresponding evolution of the energy-level 
populations for an ideal non-dissipative system (center) and a realistic dissipative 
system (right)}\label{fig:field}
\end{center}
\end{figure}

Fig.~\ref{fig:field} shows the (envelopes of the) control pulses and the corresponding
evolution of the energy-level populations if dissipative effects due to spontaneous
emission are neglected (center) and if realistic lifetimes of the excited states are
assumed (right).  The total pulse length in the example is 30 ns, where the first and 
second pulse are both 6 ns long while the third pulse is 18 ns.  The cumulative effect
of spontaneous emission is clearly noticeable by comparing the graphs.

The example studied in this paper suggests that it is possible to achieve substantial
population inversions between the ground state and a highly excited state of an atom
(or possibly ion or molecule) by applying a sequence of short control pulses, even if
the finite lifetimes of the intermediate excited states are taken into account.  In 
spite of inevitable population losses due to spontaneous emission, pulse lengths and
ratios can be adjusted to minimize dissipative effects and achieve substantial yields.
We believe such a scheme could be used to calculate the pulse configurations required
to build short wavelength pulse lasers in the UV or even X-ray regime by applying a 
sequence of short optical-frequency laser pulses to a cloud of trapped atoms or ions 
to achieve population inversion between the ground state and a highly excited state.

\end{document}